\documentstyle[prl,aps,preprint,epsf]{revtex}

\begin{document}
\title{Precursor of Non-Fermi Liquid Behaviour in the One-Dimensional Periodic
Anderson Model with Disorder}

\author{Feng Chen and Nicholas Kioussis}

\address{Department of Physics and Astronomy, California State University,
Northridge, CA 91324-8268}

\maketitle

\begin{abstract}
We have studied the one-dimensional periodic, symmetric  Anderson model 
at half filling
in the presence of disorder
using finite-temperature
quantum Monte Carlo techniques.
We have examined for the first time the disorder 
both in hybridization between
the local $f$-orbitals and the conduction electrons and in the local
$f$-site energy, 
 using a uniform 
distribution of width $\Delta$.
The $f$-orbital local magnetic moment, the uniform magnetic susceptibility,
the charge compressibility,
and the nearest-neighbor magnetic correlation function
have been calculated as a function of 
the disorder distribution width $\Delta$.
We find that the disorder in hybridization has a dramatic effect 
on the low-temperature magnetic properties exhibiting 
a non-Fermi-liquid behaviour, and that for the range
of temperature studied the  magnetic
susceptibility can be scaled by a power law with an exponent 
that is in agreement with recent experiments.
On the other hand,
disorder in the local $f$-orbital energy level
does not show a non-Fermi liquid behaviour. 
\end{abstract}                                                

\vskip 1.0 in


\newpage

The last several years have witnessed dramatic growth in both the 
experimental \cite{maple,bernal,andraka,maple2} 
and theoretical \cite{cox,kim,bulla,bhatt,dob,tesa,mira,chatt}
 activity in a new class of 
heavy fermion materials which exhibit non-Fermi-liquid (NFL) 
behaviour in their physical properties at low temperatures.
Some of the heavy fermion NFL metals, such as CeCu$_{6-x}$Au$_x$, 
have been associated with the proximity to a quantum critical 
point.\cite{lohne,millis}
  However, in several other cases of heavy-fermion 
compounds, such as Y$_{1-x}$U$_x$Pd$_3$,\cite{maple} 
UCu$_{5-x}$Pd$_x$,\cite{bernal,andraka} and 
Ce$_{1-x}$Th$_x$RhSb,\cite{andraka} 
NFL behaviour occurs 
only when the {\it f}-electron materials, consisting primarily of 
Ce or U intermetallics, have been alloyed with a 
nonmagnetic element. The NFL behaviour in these compounds is 
characterized by a linear resistivity at low T, a logarithmic low  
temperature divergence of the specific heat coefficient and 
a logarithmic or weak power law of the 
susceptibility.\cite{maple,bernal,andraka,maple2}

Several models have been proposed to explain these 
experimental results.  These include exotic single-impurity models, 
such as the quadrapolar Kondo model,\cite{cox}  various multi-channel 
Kondo models\cite{kim} and the ``compactified" Anderson impurity model.
\cite{bulla} 
Bhatt and Fischer\cite{bhatt} and Dobrosavljevi\'{c} {\it et al} 
\cite{dob} have examined the single-impurity Kondo problem 
in the presence of a random distribution of nonmagnetic impurities.
However, all these calculations strictly apply 
to systems with a {\it dilute} collection 
of Kondo centres. 
The inclusion of lattice effects in the presence of 
disorder presents an 
additional challenge that has only recently started to be 
addressed.\cite{tesa,mira,chatt}  
 Te\v{s}anovi\'{c} employed a slave boson technique.\cite{tesa} 
 More recently,
Miranda {\it et al}\cite{mira} and Chattopadhyay and Jarrell\cite{chatt} 
 have analyzed the effects of disorder 
on concentrated Kondo alloys in the limit of
 infinite spatial dimensions and of infinite U  
using the dynamical mean field theory\cite{georges}.
However, this approach does not take into account the RKKY 
interactions between the {\it f}-sites, which are 
 pertinent to the formation of the singlet ground state 
in the periodic Anderson model (PAM)\cite{scalapino}.
Another mechanism proposed as a possible cause of the NFL
behaviour is a disordered distribution of Kondo temperatures\cite{bernal}.
Castro {\it et al}\cite{castro} have attributed the NFL behaviour to the
existence of Griffiths singularities close to a quantum
critical point. These singularities arise from the
interplay between the RKKY and Kondo interactions in the
presence of magnetic anisotropy and disorder.

In this Letter, we 
 present 
the first investigation of the effects of disorder 
on the magnetic properties of the 
one-dimensional periodic Anderson model (PAM) at half filling 
 using quantum Monte Carlo techniques.   
The quantum Monte Carlo calculations treat
both the RKKY and the Kondo interactions on an equal footing, and 
allow one to treat both the weak and strong disorder regime.
We have studied the effects of disorder both in 
hybridization, V, between
the local $f$-orbitals and the conduction electrons and in the local
$f$-site energy, E$_f$,
 using a uniform
distribution of width $\Delta$.
The former case corresponds to substitution in the ligand sites with
the {\it f} sublattice remaining unchanged, whereas the latter case
corresponds to doping directly on the {\it f}-sites.
We present results for the 
 $f$-orbital local magnetic moment, the static 
uniform magnetic susceptibility,
the charge compressibility,
and the nearest-neighbor magnetic correlation function
 as a function of
the disorder distribution width $\Delta$ for both types of disorder.
The interplay of correlation and disorder can lead to 
 different types of ground states.  
In the case of disorder in hybridization, we find that 
the magnetic susceptibility exhibits 
a NFL behaviour, and that it can be scaled at low temperatures with 
a power law with an exponent that is in excellent agreement 
with experiment.\cite{maple2}
On the other hand, disorder in the local $f$-site energy leads 
to a Fermi-liquid like behaviour. 

In the absence of disorder, the PAM is 
 an interesting model since it exhibits various types 
of insulating states ranging from an antiferromagnetic insulator 
to a Kondo insulator or, if the {\it f-c} hybridization dominates, 
a simple band insulator.\cite{scalapino}
It is believed that the PAM describes the competition 
between magnetic ordering and singlet formation in a number of the 
heavy fermion systems.\cite{scalapino}
The Hamiltonian for the one-dimensional PAM is
\begin{eqnarray}
H &=&-t\sum_{i,\sigma} (c^+_{i,\sigma}c_{i+1,\sigma}+H.c.)
+E^c \sum_{i,\sigma} n^c_{i\sigma}
+\sum_{i,\sigma}V_i(f^+_{i,\sigma}c_{i,\sigma}+H.c.) \nonumber \\
& &+U_f\sum_{i} n^f_{i\uparrow} n^f_{i\downarrow}
+\sum_{i,\sigma}E^f_i n^f_{i\sigma}.
\end{eqnarray}
Here, $t$ is the hopping parameter in the $c$ band, $U_f$ is the Coulomb
repulsion on the $f$ band, $V_i$ is 
the hybridization energy between the two bands, $E^f_i$ and $E^c$
are the energy levels of the local $f$ and $c$ band, respectively, 
and $n^f_{i\sigma}
\equiv f^+_{i,\sigma}f_{i,\sigma}$ and
$n^c_{i\sigma}\equiv c^+_{i,\sigma}c_{i,\sigma}$ are the 
density operators for the 
$f$ and conduction electrons at site $i$ with spin $\sigma$.
In the following 
we will set t = 1 and consider 
 the  case at half filling 
($E^c$ = 0). 

The important thing to notice in Eq. (1) is that, unlike 
 the usual periodic Anderson model, the local {\it f}-site 
parameters V$_i$ or $E^f_i$ are taken here to be random numbers 
(static, uncorrelated disorder)
distributed according to uniform distributions 
P$_1(V_i)$ or  P$_2(E^f_i)$, respectively. Namely, 
\begin{equation}
P_1(V_i)=\frac{1}{2\Delta}\Theta(\Delta-\mid V_i-V_0 \mid ).
\end{equation}
\begin{equation}
P_2(E^f_i)=\frac{1}{2\Delta}
\Theta(\Delta-\mid E^f_i-E^f_0 \mid ).
\end{equation}
Here, $2\Delta$ denotes the width of the uniform distribution 
for each type of disorder, and 
$V_0$ and $E^f_0$ are the average values of V$_i$ and 
and E$_i^f$, respectively.
We set 
$E^f_0$ = -U$_f$/2 (symmetric case in the absence of disorder)
and  $V_0$ = U$_f$ = 1 (intermediate parameter regime).
In the intermediate coupling regime
Monte Carlo simulations provide essentially exact results,
whereas analytic approaches are most likely
to fail or are inaccurate.

We have employed a finite-temperature quantum
Monte Carlo technique with an exact updating procedure.
\cite{hirsch}\cite{gub} 
The calculated physical observables 
of the finite disordered system depend
strongly on the particular realization of the disorder.
Therefore, we have to average all quantities over a sufficient
number of disorder realizations and calculate the averaged expectation
values,
\begin{equation}
\langle \langle A \rangle \rangle_V=\int_{-\infty}^{+\infty} 
\prod dV_iP_1(V_i) \langle \hat{A} \rangle (\{V_i\}),
\end{equation}
 for the case of disorder in hybridization or 
\begin{equation}
\langle \langle A \rangle \rangle _E=\int_{-\infty}^{+\infty} 
\prod dE^f_iP_2(E^f_i)
\langle \hat{A} \rangle (\{E^f_i\}),
\end{equation}
for the case of disorder in the local E$_f$ energy level.
Here, 
$\langle \hat{A} \rangle$ denotes the thermal expectation value
of the operator $\hat{A}$
for a given disorder configuration, which  
 is calculated using the 
grand-canonical quantum Monte Carlo method developed by Fye.\cite{fye}
We have carried out simulations on an eight-site chain and 
$\beta = \frac {\large 1}{\large T} \leq 8$. We have made several checks
for sixteen-site chains. The calculated results 
for the magnetic moment, the uniform {\it f} susceptibility, 
and the correlations functions agree with those of 
the 8-site calculations  
 within the QMC statistical errors. 
This is due to the fact that the calculated observables
are local and the size effects are quite small. For the 
case of the uniform {\it f} susceptibility the main contribution
arises from the local on-site {\it f-f} contribution.

The distribution of tasks (disorder realizations) 
and the subsequent averaging procedure over disorder configurations was 
carried out in an IBM SP2 parallel machine using the Message-Passing 
Interface (MPI). We found that the statistical error is within 5\% when
the number of disorder configurations is above thirty.
Thus, the measured quantities were averaged over thirty 
different disorder configurations.

In the present study we present results of
the effect of the two different kinds of disorder 
on the following observables: (i)
The 
square of the {\it f}-orbital
 local moment $\sigma$,
\begin{equation}
\sigma =\frac{1}{N}\sum_i^N \langle \langle (n^f_{i\uparrow} - 
n^f_{i\downarrow})^2 \rangle \rangle,
\end{equation}
where $N$ is the total number of lattice sites;
(ii) The static $f$ component of the uniform magnetic susceptibility $\chi_f $,
\begin{equation}
\chi_f = \frac{1}{N} \sum_{i,j}  \int_0^\beta d\tau 
\langle \langle [n^f_{i\uparrow}(\tau) - n^f_{i\downarrow}(\tau) ]
[n^f_{j\uparrow}(0) - n^f_{j\downarrow}(0) ] \rangle \rangle ;
\end{equation}
(iii) The charge compressibility $\kappa$,
\begin{equation}
\kappa = \frac{1}{N} \sum_{i,j} \int_0^\beta d\tau \{
\langle \langle [n^f_{i\uparrow}(\tau) + n^f_{i\downarrow}(\tau) ]
[n^f_{j\uparrow}(0) + n^f_{j\downarrow}(0) ] \rangle \rangle
- \langle \langle n^f_{i\uparrow} + n^f_{i\downarrow} \rangle \rangle
\langle \langle n^f_{j\uparrow} + n^f_{j\downarrow}  \rangle \rangle \} ;
\end{equation}
and (iv) The nearest-neighbor magnetic correlation function $C(i,i+1)$,
\begin{equation}
C(i,j)=
\langle \langle [n^f_{i\uparrow} - n^f_{i\downarrow} ]
[n^f_{j\uparrow} - n^f_{j\downarrow} ] \rangle \rangle ,
\end{equation}
with $j=i+1$.

In the absence of disorder, the ground state of the one-dimensional PAM
exhibits short-range magnetic correlations;  the local
$f$-electron spin moments are compensated by correlations
with those of other $f$-electrons, as well as 
with those of the conduction electrons
leading to a nonmagnetic ground state.\cite{scalapino} 
The charge gap in the PAM is due to hybridization in the  
 absence of the Coulomb interaction  $U_f$ 
and is due to the on-site Coulomb interaction 
 when $U_f$ is larger than
the band width of the conduction electrons, since the 
{\it f}-electron decouples from the conduction electrons.\cite{guerrero}
For the parameter set used in this work 
the single-impurity Kondo temperature (T$_K \approx 0.182U
[\frac{8\Delta}{\pi U}]^{1/2} e^{-\frac{\pi U}{8\Delta}}$)
is T$_K$/t = 0.096 and 
the Kondo gap ($\Delta_K \approx T_K e^{1/2\rho J}$ )\cite{rice}
is $\Delta_K /t $ = 0.21.


We first present results for 
the case of disorder in hybridization 
which couples the  $f$-orbital and the conduction electrons.
In Fig.1 we plot the 
 impurity magnetic moment $\sigma$, the 
susceptibility $T\chi_f$, the charge compressibility $T\kappa$, and the
nearest-neighbor spin correlation function $C(i,i+1)$ as a function
of the disorder width $\Delta$ at low temperatures, $\beta$ = 8.
We find that the average of the square of 
the {\it f}-orbital local  
 magnetic moment, $\sigma$, increases as the disorder width
$\Delta$ increases. This result shows that the number 
of the {\it unquenched} local spins at low temperatures increases 
with disorder.  
Similarly, the low-temperature magnetic susceptibility $T\chi_f$,  
which  measures the ``effective" 
magnetic moment, 
  increases
with increasing disorder width.
The low value of the effective moment for $\Delta$ = 0 at low 
temperatures
indicates that the magnetic correlations act to screen the 
{\it f}-site moment leading to a singlet ground state.
Turning on the disorder in hybridization results in 
a destruction of the singlet ground-state and a reduction 
 of the spin gap between the singlet ground state and 
the lowest lying excited triplet state.\cite{guerrero} 
Fig. 1 shows also that 
the nearest-neighbor spin correlation function $C(i,i+1)$ 
increases monotonically with $\Delta$.
These short-range {\it f - f} 
antiferromagnetic correlations 
have been found to be as important as the {\it f - c}
correlations in the formation of the singlet state in 
the PAM.\cite{scalapino,fye}
Consistent with the above results, we find that
 increase of disorder in hybridization tends to 
suppress these short-range antiferromagnetic 
{\it f - f} correlations, resulting 
in non-compensated {\it f} moments.
On the other hand, 
 the charge compressibility
$T\kappa$, shown also in Fig. 1,  
is almost unchanged by disorder in hybridization.    
 This suggests that
the charge gap might be due to correlation effects rather than 
 hybridization
effects.


In order to 
gain further insight into the 
effect of the hybridization-disorder  
on the magnetic properties of the $f$-electrons,
 we present in Fig. 2 the {\it f} magnetic susceptibility $\chi_f$
as a function of temperature for 
 $\Delta = 0$ (PAM) and $\Delta = 1$. 
 In agreement with 
a previous study for the PAM ($\Delta = 0$),\cite{fye} we find 
 that as the 
temperature is lowered  $\chi_f$ saturates to 
a constant value. On the other hand, the {\it total} 
uniform magnetic susceptibility which contains
both the {\it f} and the {\it conduction electron} components
goes to zero due to the presence of the spin gap.\cite{fye2}
This Fermi liquid 
behaviour in the absence of disorder signals  
the compensation of the {\it f} moment by the 
antiferromagnetic {\it f - f} spin correlations in addition to the  
already-present Kondo-like fluctuations from the 
antiferromagnetic {\it c - f} screening correlations.\cite{scalapino,fye}
On the other hand, in the presence of disorder, 
$\chi_f$ exhibits a non-Fermi-liquid behaviour, 
 divering 
at low temperatures. 
We find that $\chi_f$ can 
be scaled at low temperatures by a power law, 
$\chi_f \sim T^{-\gamma}$,  
 with the exponent $\gamma = 0.66$, which is comparable
to recent experimental results.\cite{maple2}
This result might be interpreted that the hybridization-disorder 
causes some of the 
local {\it f} sites to possess  
low Kondo temperatures and consequently 
to behave as nearly free magnetic moments, leading to  
a NFL behaviour.


We next present results for the effect of disorder in the 
 $f$-electron energy level, {\it E$_f$}, on the magnetic properties.
 In Fig. 3 we show the impurity local magnetic
moment $\sigma$, the magnetic susceptibility $T\chi_f$, 
the charge compressibility
$T\kappa$, and the nearest-neighbor {\it f - f} 
spin correlation function $C(i,i+1)$
as a function of the {\it f}-energy-level disorder width $\Delta$ 
at temperature $T/t = 1/8$.
Note, that the range of $\Delta$ values in Fig. 3 is larger than that
 in Fig. 1.
In the weak-disorder regime, 
  $\Delta/t \leq 1$, all quantities 
are almost unchanged by the disorder in E$_f$,
 in contrast to the corresponding results shown 
in Fig. 1 for the case of hybridization in V.
This is due to the fact that the 
majority of the {\it f} sites are 
occupied, since the average 
$f$-electron
energy level is below the Fermi energy ($E^f_0 = -U/2$). 
On the other hand, in the strong-disorder regime, $\Delta/t > 1$,
 some of the local {\it f}-energy levels are distributed 
above the Fermi energy as $\Delta$ increases further. 
This in turn causes the average local magnetic moment $\sigma$ to
decrease with $\Delta$ because some of the 
{\it f}-orbitals become empty. 
The ``effective moment"  
$T\chi_f$ increases slightly with $\Delta$, indicating  
that even though a few {\it diluted} {\it f} spins 
cease to contribute 
to the compensation state,
 the singlet ground state is not destroyed. 
The nearest-neighbor
spin correlation function $C(i,i+1)$ is suppressed because 
some of the effective
$f$-electron orbitals are diluted by disorder. 
The low-temperature charge compressibility $\kappa$
increases with the disorder width 
in the strong-disorder regime. 
Strong disorder in the distribution of the $E_f$'s leads 
to a reduction of the charge gap and hence an increase of $\kappa$.

In conclusion, we have studied the effect of 
two kinds of disorder on the magnetic properties of the 
one-dimensional PAM using quantum Monte Carlo
simulations. 
We find that disorder in 
hybridization  modifies considerably  
 the low-temperature properties  
due to the presence of 
 {\it unquenched} local moments. 
The low-temperature susceptibilty exhibits a NFL hehaviour
which is found to diverge with 
a power law with an exponent  in agreement 
with experiment.\cite{maple2} 
On the other hand, disorder in the {\it f}-site energy level
results in a conventional Fermi liquid behaviour.

\acknowledgments{The research at California State University Northridge (CSUN)
was supported through the National Science Foundation under Grant
No. DMR-9531005, the Parsosn Foundation 
and the Office of Research and Sponsored Projects
at CSUN.}

\newpage
\begin{figure}
\caption{The square of the {\it f}-orbital 
 local moment $\sigma$ (open squares), 
the {\it f} magnetic susceptibility $T\chi_f$ (solid squares),
the charge susceptibility $T\kappa$ (sold triangles) and 
the nearest-neighbor {\it f - f} spin correlation function $C(i,i+1)$
(open triangles)
as a function of the hybridization-disorder width $\Delta$ 
at temperature $T/t=1/8$
with $U_f/t = 1$ and $V/t =1$.}
\epsfxsize = 7.0 in
\epsffile{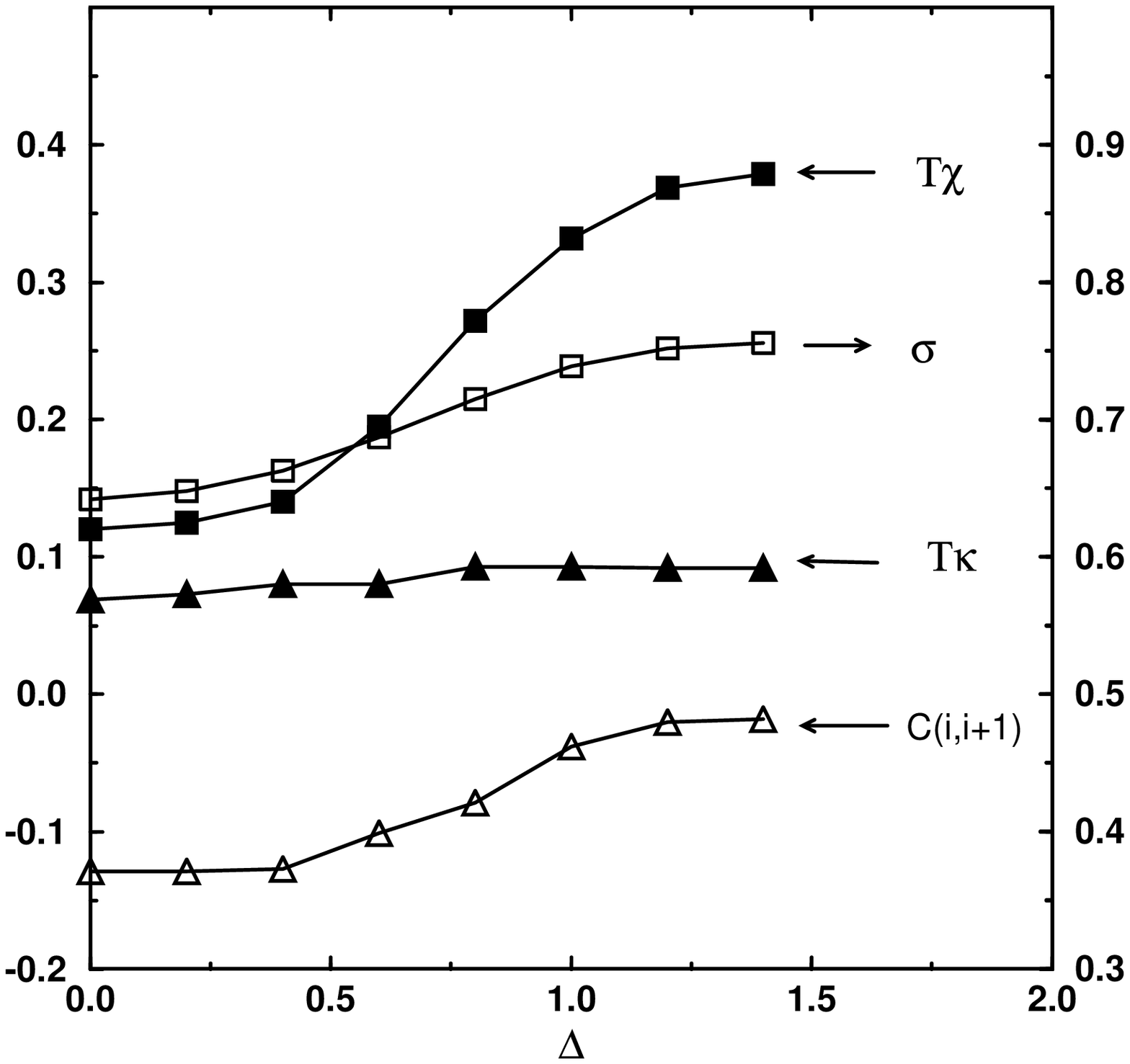}
\end{figure}

\newpage
\begin{figure}
\caption {The uniform {\it f} magnetic susceptibility as a function 
of temperature for the hybridization-disorder width 
$\Delta = 0$(triangles) and $\Delta = 1$
(circles) for $U_f/t = 1$ and $V/t =1$.
For ($T/t \leq 1$), $\chi_f$ 
can be scaled by the power law, $ \chi_f \sim
T^{-\gamma}$, with $\gamma$ = 0.66.
}
\vskip 0.5 in
\epsfxsize = 7.0 in
\epsffile{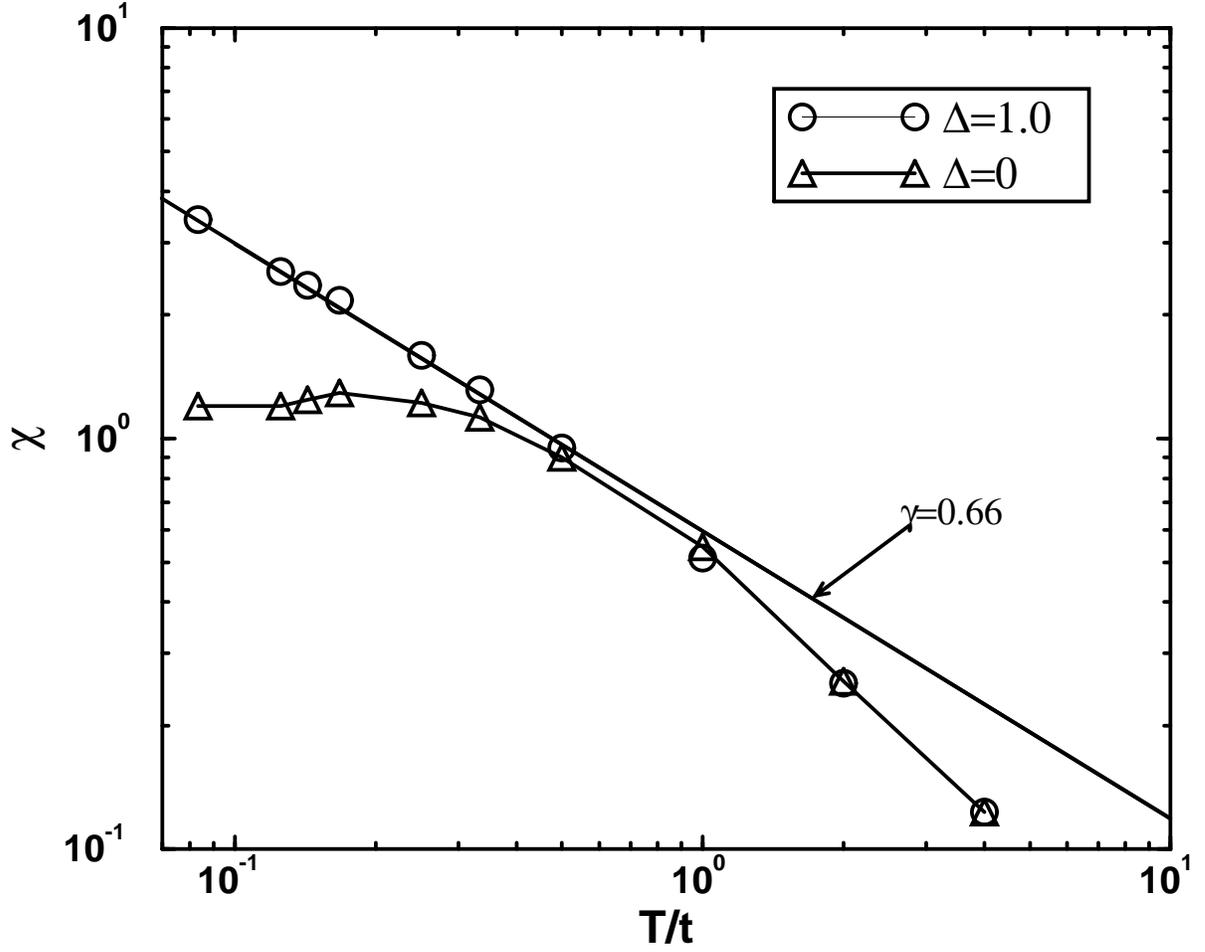}
\end{figure}

\newpage
\begin{figure}
\caption{The square of the {\it f}-orbital
 local moment $\sigma$ (open squares),
the {\it f} magnetic susceptibility $T\chi_f$ (solid squares),
the charge susceptibility $T\kappa$ (sold triangles) and
the nearest-neighbor {\it f - f} spin correlation function $C(i,i+1)$
(open triangles)
as a function
 of the {\it f}-energy-level disorder width 
$\Delta$ 
at temperature $T/t=1/8$
with $U_f/t = 1$ and $V/t =1$.}

\epsfxsize = 7.0 in
\epsffile{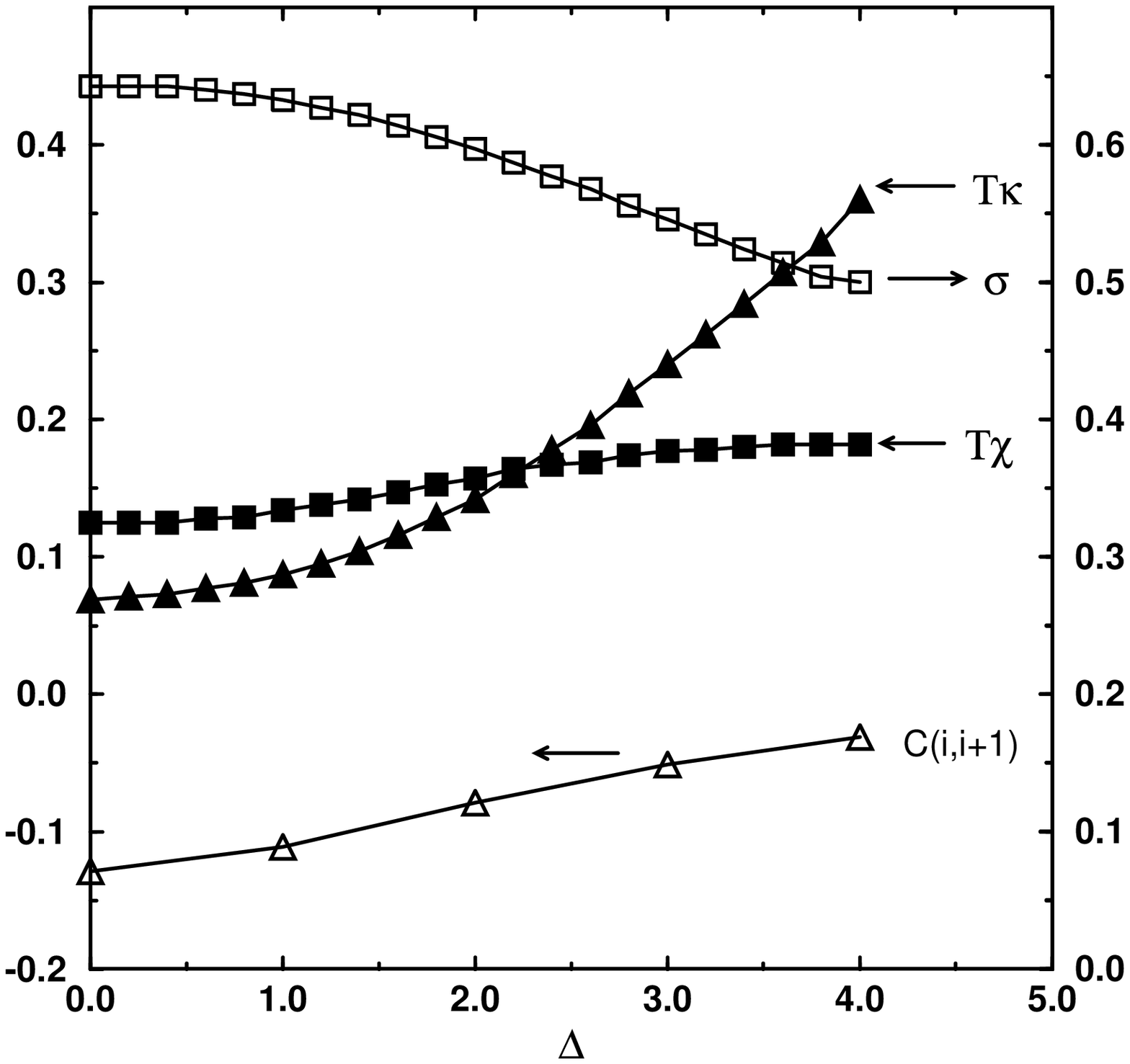}
\end{figure}


\newpage
\begin{references}

\bibitem{maple} M.B. Maple {\it et al.}, 1995, {\it J. Low Temp. Phys.},
{\bf 99,} 223.

\bibitem{bernal} O. O. Bernal, D. E. MacLaughlin, H. G. Lukefahr,
and B. Andraka, 1995, {\it Phys. Rev. Lett.}, {\bf 75,} 2023.

\bibitem{andraka} B. Andraka and G. R. Stewart, 1993, {\it
Phys. Rev. B}, {\bf 47,} 
3208 (1993); B. Andraka, 1994, {\it Phys. Rev. B}, {\bf 49,} 348.

\bibitem{maple2} M. C. de Andrade {\it et al.}, 1998, {\it cond-mat 
9802081}.

\bibitem{cox} D. L. Cox, 1987, {\it Phys. rev. Lett.}, {\bf 59,} 1240.

\bibitem{kim} T.S. Kim and D. L. Cox, 1995, {\it Phys. Rev. Lett.}, {\bf 75,} 
1622; P. Nozi\`{e}res and P. Blandin, 1980, {\it J. Physique} (Paris), 
{\bf 41,} 193.

\bibitem{bulla} R. Bulla, A. C. Hewson, 1997,
{\it Physica B}, {\bf 230-232,} 627.

\bibitem{bhatt} R. N. Bhatt and D. S. Fisher, 1992, {\it Phys. Rev. Lett.,}
{\bf 68,} 3072.

\bibitem{dob} V. Dobrosavljevi\'{c}, T. R. Kirkpatrick, and G. Kotliar, 
1992, {\it Phys. Rev. Lett.,} {\bf 69,} 1113.

\bibitem{tesa} 
Z. Te\v{s}anovi\'{c}, 1986, {\it Phys. Rev. B,} {\bf 34,} 5212.

\bibitem{mira} E. Miranda, V. Dobrosavljevi\'{c} and G. Kotliar,
1997, {\it Phys. Rev. Lett.}, {\bf 78,} 290.

\bibitem{chatt} A. Chattopadhyay and M. Jarrell, 1997, {\it Phys. Rev. B},
{\bf 56,} 2920.

\bibitem{lohne}H. L\"{o}hneysen {\it et al}, 1994, {\it Phys. Rev. Lett. },
{\bf 72,} 3262; B. Bogenberger and H. L\"{o}hneysen, 
1995, {\it Phys. Rev. Lett.}, {\bf 74,} 1016.

\bibitem{millis} M. A. Continentino, 1993, {\it Phys. Rev. B }, {\bf 47,} 
11587; A. J. Millis, 1993, {\it Phys. Rev. B}, {\bf 48,} 7183.

\bibitem{georges} A. Georges, G. Kotliar and Q. Si, 
1992, {\it Intern. J. Mod. Phys. B}, {\bf 6,} 705. 

\bibitem{scalapino} R. Blankenbecler, J. R. Fulco, W. Gill, 
and D. J. Scalapino, 1987, {\it Phys. Rev. Lett.}, {\bf 58,} 411.

\bibitem{castro} A. H. Castro Neto, G. Castilla, and B. A. Jones, 1998,
{\it Phys. Rev. Lett.}, {\bf 81,} 3531.

\bibitem{hirsch} J. E. Hirsch, and R. M. Fye, 1986, {\it Phys. Rev. Lett.},
{\bf 56,} 2521.

\bibitem{gub} J. E. Gubernatis, J. E. Hirsch, and D. J. Scalapino, 
1987, {\it Phys. Rev.  B}, {\bf 35,} 8478 .

\bibitem{fye} R. M. Fye, 1990, {\it Phys. Rev. B}, {\bf 41,} 2490.

\bibitem{guerrero}M. Guerrero and C. C. Yu, 1995, {\it Phys. Rev. B}, {\bf 51,}
10301.

\bibitem{rice} T. M. Rice and K. Ueda, 1986,
{\it Phys. Rev. B}, {\bf 34,} 6420.

\bibitem{fye2} R. M. Fye and D. J. Scalapino, 1990, {\it Phys. Rev. Lett.},
{\bf 65,} 3177.

\end{references}
\end{document}